\documentclass[a4paper]{article}

\usepackage{arxiv}

\usepackage[utf8]{inputenc} 
\usepackage[T1]{fontenc}    
\usepackage{hyperref}       
\usepackage{url}            
\usepackage{booktabs}       
\usepackage{amsfonts}       
\usepackage{nicefrac}       
\usepackage{microtype}      
\usepackage{lipsum}		
\usepackage{graphicx}
\usepackage{natbib}
\usepackage{doi}
\usepackage{latexsym}
\usepackage{amsmath}
\usepackage{amssymb}
\usepackage[mathscr]{eucal}
\usepackage{graphicx,color}
\usepackage{mathtools}
\usepackage{stmaryrd}
\usepackage{url}
\usepackage{xypic}

\usepackage{bm}

\providecommand{\R}{\mathbb{R}}

\providecommand{\SO}{\mathbf{SO}}

\usepackage{accents}
\makeatletter
\providecommand{\scirc}{%
    \hbox{\fontfamily{\rmdefault}\fontsize{0.4\dimexpr(\f@size pt)}{0}\selectfont{\raisebox{-0.52ex}[0ex][-0.52ex]{$\circ$}}}}

\makeatother

\mathchardef\mhyphen="2D

\usepackage{bbm}
\usepackage{tikz-cd}
\usepackage{psfrag}

\providecommand{\R}{\mathbb{R}}
\providecommand{\Sp}{\mathrm{S}}
\providecommand{\SO}{\mathbf{SO}}

\usepackage{amsthm}
\newtheorem{theorem}{Theorem}

\newtheorem{definition}{Definition}
\newtheorem{assumption}{Assumption}
\newtheorem{lemma}{Lemma}
\newtheorem{remark}{Remark}

\title{
Vision-Aided Relative State Estimation for Approach and Landing on a Moving Platform with Inertial Measurements}

    \author{ \href{https://orcid.org/0000-0003-1116-7415}{\includegraphics[scale=0.06]{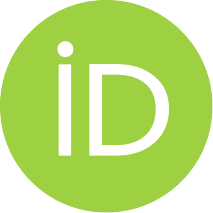}\hspace{1mm}Tarek Bouazza} \\
	I3S, CNRS, Université Côte d'Azur\\
        06900 Sophia Antipolis, France \\
	\texttt{bouazza@i3s.unice.fr} \\
     \And
     \href{https://orcid.org/0009-0005-8795-8186}{\includegraphics[scale=0.06]{orcid.pdf}\hspace{1mm}Alessandro Melis} \\ 
	I3S, CNRS, Université Côte d'Azur\\
        06900 Sophia Antipolis, France \\
	\texttt{melis@i3s.unice.fr} \\
	\And
     \href{https://orcid.org/0000-0002-8718-8004}{\includegraphics[scale=0.06]{orcid.pdf}\hspace{1mm}Soulaimane Berkane} \\
        Département d’informatique et d’ingénierie, \\
	Université du Québec en Outaouis \\
        and Department of Electrical Engineering, \\ 
        Lakehead University \\
        Gatineau, QC J8X 3X7, Canada \\
	\texttt{soulaimane.berkane@uqo.ca} \\
	\And
    \href{https://orcid.org/0000-0002-7803-2868}{\includegraphics[scale=0.06]{orcid.pdf}\hspace{1mm}Robert Mahony} \\
	Systems Theory and Robotics Group\\
        Australian National University\\
	ACT, 2601, Australia \\
	\texttt{Robert.Mahony@anu.edu.au} \\
     \And
	\href{https://orcid.org/0000-0002-7779-1264}{\includegraphics[scale=0.06]{orcid.pdf}\hspace{1mm}Tarek Hamel} \\
    I3S, CNRS, Université Côte d'Azur\\
    and Institut Universitaire de France (IUF) \\
    06900 Sophia Antipolis, France \\
	\texttt{thamel@i3s.unice.fr} \\
}

\renewcommand{\headeright}{}
\renewcommand{\undertitle}{Author accepted version\thanks{
\, \textcopyright 2026 the authors. This work has been accepted to IFAC for publication under a Creative Commons Licence CC-BY-NC-ND.
T. Bouazza, A. Melis, S. Berkane, R. Mahony and T. Hamel, ``Vision-Aided Relative State Estimation for Approach and Landing on a Moving Platform with Inertial Measurements', in Proceedings
of IFAC World Congress, 2026.
}}

\renewcommand{\shorttitle}{}

\hypersetup{
pdftitle={A template for the arxiv style},
pdfsubject={q-bio.NC, q-bio.QM},
pdfauthor={David S.~Hippocampus, Elias D.~Striatum},
pdfkeywords={First keyword, Second keyword, More},
}

\begin{document}
\maketitle

\begin{abstract}
This paper tackles the problem of estimating the relative position, orientation, and velocity between a UAV and a planar platform undergoing arbitrary 3D motion during approach and landing. 
The estimation relies on measurements from Inertial Measurement Units (IMUs) mounted on both systems, assuming there is a suitable communication channel to exchange data, together with visual information provided by an onboard monocular camera, from which the bearing (line-of-sight direction) to the platform’s center and the normal vector of its planar surface are extracted.
We propose a cascade observer with a complementary filter on $\SO(3)$ to reconstruct the relative attitude, followed by a Riccati observer for relative position and velocity estimation. Convergence of both observers is established under persistently exciting conditions, and the cascade is shown to be almost globally asymptotically and locally exponentially stable. We further extend the design to the case where the platform’s rotation is restricted to its normal axis and show that its measured linear acceleration can be exploited to recover the remaining unobservable rotation angle. A sufficient condition for local exponential convergence in this setting is provided. The proposed observers are validated through extensive simulations.
\end{abstract}

\noindent\textbf{Keywords:}
    Nonlinear observers and filter design, Estimation and filtering, Information and sensor fusion, UAVs, Mobile robots.

\section{Introduction}
Autonomous landing of Unmanned Aerial Vehicles (UAVs) has been a popular research topic for over two decades.
Early work primarily focused on static targets such as runways or landing pads. 
The UAV state was typically estimated using Global Positioning Systems (GPS), possibly augmented by onboard cameras that detect fiducial markers or structured visual features to determine the relative position of the landing target. 
However, GPS signals are unreliable in many relevant scenarios and even when available they are imprecise and subject to time-lag. 
The question of precision can be addressed using real-time kinematic GPS (GPS-RTK) \citep{kang2018precision} if such a system is available. 
Alternatively, motion-capture systems \citep{mellinger2010control} have been used to provide relative state estimation, however, such infrastructure is expensive, especially for landing on moving targets in outdoor environments. 
As a consequence, most research in approach and landing for moving targets has exploited vision-based techniques, where the UAV detects and tracks the platform using an onboard camera \citep{falanga2017vision}.
Visual servoing techniques are frequently used to control the UAV using image feedback \citep{lee2012autonomous,serra2016landing}.
To improve robustness, some approaches equip the platform with additional sensors, such as infrared markers \cite{Wenzel2011Automatic}.
When the target undergoes arbitrary motion, estimating the relative states becomes significantly more challenging, as the platform’s dynamics must be explicitly accounted for in the estimator. 
Cooperative strategies, in which the UAV leverages measurements from ground or marine vehicles to assist in relative motion estimation, have been explored in \citep{Borowczyk2017Autonomous, Acuna2018Vision, Cui2023Coarse}.
While effective, these methods implicitly constrain the platform’s motion to the kinematic models used in their filters and do not address the general free motion case.

In this work, we consider the estimation problem for the approach and landing of a UAV on a moving target. 
The goal is to recover the relative pose and velocity of the UAV with respect to the target using IMU measurements from both systems and monocular vision. 
We assume that the vision system provides the bearing to the landing platform’s center and the normal to the platform surface.
We propose a cascade observer architecture that first estimates the relative attitude through a complementary filter on $\SO(3)$, and subsequently reconstructs the relative position and velocity using a linear Riccati observer. Convergence is established under persistently exciting conditions, and the cascade interconnection is shown to achieve almost global asymptotic and local exponential stability. An extension is developed for the case where the target’s rotation is restricted to its normal axis, which exploits its linear acceleration to estimate the yaw rotation angle and derive sufficient conditions for local exponential convergence.

The remainder of the paper is organized as follows. Section~\ref{sec:notation} introduces notation and preliminaries. Section~\ref{sec:problem_description} formulates the problem with the available inertial and visual measurements and derives relative UAV-target dynamics. Section \ref{sec:cascade_observer} develops the cascade observer, first under persistent excitation of the target-plane normal, and then extends the design to the case where this excitation condition is not satisfied.
Simulation results are reported in Section~\ref{sec:simulations}, and Section~\ref{sec:conclusion} concludes the paper. Proofs of the lemmas are provided in the Appendix.

\section{Preliminary Material} \label{sec:notation}
\subsection{Notation}

The $n$-dimensional Euclidean space is denoted by $\R^n$.
The Euclidean norm of a vector $\bm{x} = (x_1, \dots, x_n) \in \mathbb{R}^n$ is $\|\bm{x}\| := \sqrt{\sum_1^n x_i^2}$.
The unit $n$-sphere $\mathrm{S}^n := \{\bm{u} \in \mathbb{R}^{n+1} \mid \|\bm{u}\| = 1\}$ denotes the set of unit vectors in $\R^{n+1}$.
For any $\bm{u} \in\mathrm{S}^{2}$, the projection operator $\pi_{\bm{u}} :=I_{3}-\bm{u} \bm{u}^\top$ projects any $\bm{x} \in \R^3$ to the plane orthogonal to $\bm{u}$.

We denote by $\mathrm{R}^{m \times n}$ the set of real $m \times n$ matrices. The set of $n \times n$ positive definite matrices is denoted by $\mathbb{S}_+(n)$, and the identity matrix is denoted by $I_n \in \mathrm{R}^{n \times n}$.
    
The special orthogonal group of 3D rotations is given by
$\SO(3) := \{R \in \mathbb{R}^{3\times3} \mid R^{\top} R = RR^\top = I_3, \det(R) = 1\}$.
Its Lie algebra is $\mathfrak{so}(3) := \{\bm{\omega}^\times \in \mathbb{R}^{3\times3} \mid \bm{\omega} \in \mathbb{R}^3\}$,
where $(\cdot)^\times: \mathbb{R}^3 \to \mathfrak{so}(3)$ denotes the skew-symmetric matrix operator associated with the cross product, which satisfies $\bm{\omega}^\times \bm{v}=\bm{\omega}\times\bm{v}$, for all $\bm{\omega},\bm{v}\in\mathbb{R}^3$. 

The exponential map $\exp : \mathfrak{so}(3) \rightarrow \SO(3)$ defines a local diffeomorphism from a neighbourhood of $0 \in \mathfrak{so}(3)$ to a neighbourhood of $I_3 \in \SO(3)$.
It allows us to define the composition map $\exp \circ (\cdot)^\times: \mathbb{R}^3 \rightarrow \SO(3)$, 
which provides a convenient parametrization of rotations. 
For any vector $\theta \bm{u} \in \R^3$, with angle $\theta$ and axis $\bm{u} \in \Sp^2$, this map can be computed explicitly via Rodrigues’ formula: 
\begin{equation} \label{eq:rodrigues_formula}
\exp(\theta \bm{u}^\times) = I_3 + \sin(\theta)\bm{u}^\times + (1 - \cos(\theta)) \bm{u}^\times \bm{u}^\times. 
\end{equation}

\subsection{Stability and Observability Definitions} 
Consider a generic linear time-varying (LTV) system:
\begin{subequations} \label{eq:LTV_system_dynamics}
    \begin{align}
        \dot{\bm{x}} &= A(t) \bm{x} + B(t) \bm{u} \label{eq:state_equation}, \\
       \bm{y} &=C(t) \bm{x} \label{eq:measurement_equation},
    \end{align}
\end{subequations}
where \( \bm{x} \in \mathbb{R}^n \) denotes the system state, \( \bm{u} \in \mathbb{R}^s \) the system input, and \( \bm{y} \in \mathbb{R}^m \) the system output. The following definition of observability for this system is borrowed from \citep{chen1984linear}.
\begin{definition}[Uniform Observability] \label{def:uniform_obs}
The system \eqref{eq:LTV_system_dynamics} is said to be \textit{uniformly observable} if there exist constants $\delta, \mu > 0$ such that, for all $t \geq 0$:  
\begin{equation}  
W(t, t + \delta) \succeq \mu I_n \succ 0, 
\label{eq:main Uniform Observability condition}
\end{equation}  
where  
\begin{equation} \label{eq:observability_gramian}
W(t, t + \delta) := \frac{1}{\delta} \int_{t}^{t+\delta} \Phi^\top(s, t)C^\top(s)C(s)\Phi(s, t) \, ds 
\end{equation} is the observability Gramian, and  
$\Phi(t, s)$ denotes the state transition matrix associated with $A(t)$, defined by: $\frac{d}{dt} \Phi(t, s) = A(t) \Phi(t, s)$, $\Phi(t, t) = I_n$.
If condition \eqref{eq:main Uniform Observability condition} holds, the pair $(A(t), C(t))$ is said to be \textit{uniformly observable}.
\end{definition}

\begin{definition}[Almost global asymptotic stability]
    An equilibrium point of a nonlinear system is said to be almost globally asymptotically stable if it is locally stable and its basin of attraction contains all initial conditions except for a set of measure zero.
\end{definition}

\section{Problem description} \label{sec:problem_description}

This work aims to design an observer that estimates the robot’s relative pose and velocity with respect to a moving landing platform.
Both the robot and the platform are equipped with onboard IMUs that provide their angular velocity and linear acceleration measurements. While these measurements capture their individual dynamics, the relative states are required to perform a successful landing. 

\begin{figure}[h]
    \centering
    \includegraphics[width=0.65\linewidth, trim=0 0 0 10, clip]{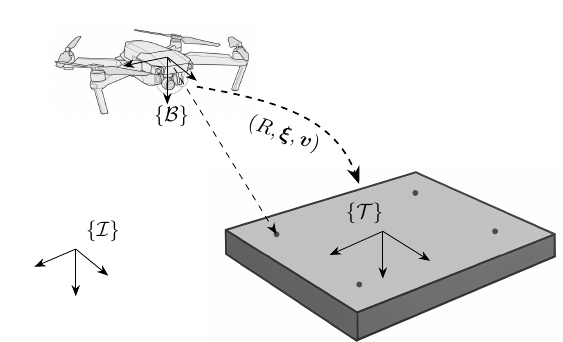}
    \caption{Illustration of the relative extended pose $(R, \bm{\xi}, \bm{v})$ estimation problem, where both the UAV frame $\{\mathcal{B}\}$ and the target frame $\{\mathcal{T}\}$ are moving with respect to the inertial frame $\{\mathcal{I}\}$. The target carries visual landmarks observed by an onboard monocular camera.}
    \label{fig:uav_target_setup}
\end{figure}

Let $\{\mathcal{B}\}, \{\mathcal{T}\}$ and $\{\mathcal{I}\}$ denote, respectively, the UAV body frame, the target (platform) frame, and an inertial reference frame (see Fig.~\ref{fig:uav_target_setup}). The orientation of frame $\{\mathcal{B}\}$ (resp. $\{\mathcal{T}\}$) w.r.t. frame $\{\mathcal{I}\}$ is represented by the rotation matrix $Q_B \in \mathbf{SO}(3)$ (resp. $Q_T \in \mathbf{SO}(3)$). 
Denote as $\bm{p}_B\in \mathbb{R}^3$ (resp. $\bm{p}_T\in \mathbb{R}^3$) and $\bm{v}_B\in\mathbb{R}^3$ (resp. $\bm{v}_T\in \mathbb{R}^3$) the position and linear velocity of the body (resp. target) expressed in the inertial frame $\mathcal{I}$.

We assume that both the UAV and the platform are equipped with an IMU. Let $\bm{\omega}_B,\bm{\omega}_T\in\mathbb{R}^3$ denote the angular velocities of the body and target, expressed in frames $\mathcal{B}$ and $\mathcal{T}$, respectively, and let $\bm{a}_B,\bm{a}_T\in\mathbb{R}^3$ denote the corresponding specific accelerations measured by their accelerometers. Their rigid-body motion then evolves according to
\begin{equation}\label{pose kinematics}
     \begin{cases}
        \dot{Q}_B = Q_B \bm{\omega}_B^\times\\
        \dot{\bm{p}}_B = \bm{v}_B \\
        \dot{\bm{v}}_B = Q_B \bm{a}_B + g \bm{e}_3,
    \end{cases} \quad 
    \begin{cases}
        \dot{Q}_T = Q_T \bm{\omega}_T^\times \\
        \dot{\bm{p}}_T = \bm{v}_T \\
        \dot{\bm{v}}_T = Q_T \bm{a}_T + g \bm{e}_3,
    \end{cases}
\end{equation}
where $g \approx 9.81m/s^2$ the gravity constant and $\bm{e}_3 = (0,0,1)^\top$ the vertical direction.

The position of frame $\{\mathcal{B}\}$ w.r.t. frame $\{\mathcal{T}\}$ expressed in the frame $\{\mathcal{B}\}$ 
is represented by  $\bm{\xi} \in \mathbb{R}^3$, such that 
\begin{align} \label{p_xi}
    \bm{\xi} = Q_B^\top ( \bm{p}_B - \bm{p}_T).
\end{align}
From \eqref{p_xi}, the relative linear position evolution is
\begin{align} \label{eq:dyn_p}
    \dot{\bm{\xi}} 
    &= -\bm{\omega}_B^\times Q_B^\top ( \bm{p}_B - \bm{p}_T) + Q_B^\top ( \dot{\bm{p}}_B - \dot{\bm{p}}_T) \notag\\
    &= -\bm{\omega}_B^\times Q_B^\top ( \bm{p}_B - \bm{p}_T) + Q_B^\top ( \bm{v}_B - \bm{v}_T) \notag\\
    &= -\bm{\omega}_B^\times \bm{\xi} + \bm{v},
\end{align}
where $\bm{v} := Q_B^\top (\bm{v}_B - \bm{v}_T)$ represents the relative body-target linear velocity expressed in frame $\mathcal{B}$.
From \eqref{pose kinematics}, the relative linear velocity dynamics are given by
\begin{align} \label{eq:dyn_v}
    \dot{\bm{v}} & = -\bm{\omega}_B^\times Q_B^\top(\bm{v}_B - \bm{v}_T) + Q_B^\top (\dot{\bm{v}}_B - \dot{\bm{v}}_T) \notag \\
    &= -\bm{\omega}_B^\times \bm{v} + Q_B^\top(Q_B \bm{a}_B + g \bm{e}_3 - Q_T \bm{a}_T - g \bm{e}_3) \notag \\
    &= -\bm{\omega}_B^\times \bm{v} + \bm{a}_B - R^\top \bm{a}_T.
\end{align}
where $R := Q_T^\top Q_B \in \mathbf{SO}(3)$ denotes the orientation of frame $\mathcal{B}$ with respect to frame $\mathcal{T}$, with dynamics  
\begin{align} \label{eq:R_dynamics}
\dot R = - \bm{\omega}_T^\times R + R\bm{\omega}_B^\times . 
\end{align}

We assume the UAV is equipped with a monocular camera observing at least four landmarks rigidly attached to the target (see Fig.~\ref{fig:uav_target_setup}), such as ArUco or AprilTag corners. 
Given these landmarks, a P$n$P algorithm \citep{lepetit2009ep} can be used to extract relative pose information. In particular, we adopt the hierarchical measurement structure proposed in \citep{bouazza2025simple}, which yields three key pieces of information recovered hierarchically:

\begin{enumerate}
    \item \textbf{Bearing vector:} the unit vector pointing from the target center to the camera,
    \begin{equation} \label{eq:bearing}
        \bm{y}_\xi := \frac{\bm{\xi}}{\|\bm{\xi}\|} \in \Sp^2,
    \end{equation}
    can be reliably estimated even when the camera is far from the target, providing robust long-range direction information.
    
    \item \textbf{Normal to the target plane:} once the target appears clearly in the image, the unit vector normal to the target plane can be recovered from one of several sets of four landmarks,
    \begin{equation} \label{eq:eta_output}
        \bm{\eta} := R^\top \bm{e}_3 \in \Sp^2,
    \end{equation}
    where $\bm{e}_3$ is the vertical direction in $\mathcal{T}$. 
    
    \item \textbf{Full orientation and scale:} the remaining degrees of freedom (orientation and scale) can be recovered once sufficient geometric information is available.
\end{enumerate}

A key advantage of this hierarchical approach is that partial pose information (e.g., the bearing vector) can be accurately recovered even when the UAV is far from the target. In contrast, the normal vector and the remaining components are estimated only when the target is clearly visible. To ensure robust estimation across a wide range of distances, we formulate the estimation problem using only the bearing \eqref{eq:bearing} and the target's normal direction \eqref{eq:eta_output}.

\section{Cascade observer design} \label{sec:cascade_observer}
\begin{figure}[ht]
    \centering
    \includegraphics[width=0.7\linewidth]{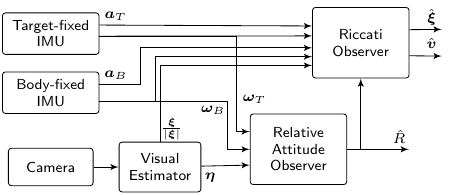}
    \caption{Diagram of the cascade observer architecture. The relative attitude $R$ is estimated via a complementary filter on $\mathbf{SO}(3)$ using the normal measurement $\bm{\eta}$, and then together with the bearing and UAV-target IMU measurements, is used in a Riccati observer to reconstruct the relative position and velocity $(\bm{\xi},\bm{v})$.}
    \label{fig:cascade_architecture}
\end{figure}

We proceed by designing an observer for the relative attitude dynamics of $R$. The resulting estimate is then injected, in a cascade fashion, into the $(\bm{\xi},\bm{v})$-subsystem. The latter becomes an LTV system parametrized by the estimate $\hat{R} \in \SO(3)$, for which we subsequently construct a linear Riccati observer (see Fig.~\ref{fig:cascade_architecture}).

\subsection{Attitude observer}

We define the relative attitude estimate $\hat{R} \in \SO(3)$ for $R$,
and propose an observer of the form
\begin{align} \label{R_observer}
        \dot{\hat{R}} &=  - \bm{\omega}_T^\times \hat{R} + \hat{R} \bm{\omega}_B^\times + \bm{\sigma}_R^\times \hat{R}, & \hat{R}(0) &= \hat{R}_0, 
\end{align}
where $\bm{\sigma}_{R}\in \mathbb{R}^3$ denotes an innovation term to be designed to ensure convergence of $\hat{R}$ to $R$. 

Consider the classical right-invariant error $\tilde{R} := \hat{R} R^\top$ on $\SO(3)$. Using \eqref{eq:R_dynamics} and \eqref{R_observer}, one verifies that $\tilde{R}$ satisfies 
\begin{align} \label{eq:Rtilde_dynamics}
    \dot{\tilde{R}} &=  (- \bm{\omega}_T^\times \hat{R} + \hat{R} \bm{\omega}_B^\times  + \bm{\sigma}_R^\times \hat{R} ) R^\top + \hat{R} (R^\top \bm{\omega}_T^\times -\bm{\omega}_B^\times R^\top) \notag \\
    &= \tilde{R} \bm{\omega}_T^\times - \bm{\omega}_T^\times \tilde{R}  + \bm{\sigma}_R^\times \tilde{R}  \notag \\
    &= [\tilde{R}, \bm{\omega}_T^\times ]  + \bm{\sigma}_R^\times \tilde{R}.
\end{align}  
where $[\cdot,\cdot]$ is the matrix commutator (Lie bracket). 

\begin{lemma} \label{lemma_attitude_observer}
Consider the attitude error dynamics \eqref{eq:Rtilde_dynamics} with measurement \eqref{eq:eta_output} and innovation term
\begin{align}\label{sigma_R}
\bm{\sigma}_{R} &= 2k_R ( (\hat{R} \bm{\eta}) \times \bm{e}_3), & k_R &> 0.
\end{align}
Assume that $\bm{\omega}_B$ and $\bm{\omega}_T$ are bounded and uniformly continuous,
and that the 
    normal direction to the target expressed in the inertial frame $\bm{\eta}_\mathcal{I} := Q_T \bm{e}_3 \in \Sp^2$
    is persistently exciting. That is, there exist $\mu, \delta >0$ such that for all $t \geq 0$,
\begin{align}\label{eq:PE eta}
\int_t^{t+\delta} \pi_{\bm{\eta}_\mathcal{I}(s)} ds = \int_t^{t+\delta}(I_3 - \bm{\eta}_\mathcal{I} (s)\bm{\eta}_\mathcal{I} (s)^\top) ds \succeq \mu I_3.
\end{align}
Then, the error $\tilde{R}$ is almost globally asymptotically and locally exponentially stable at the identity $I_3$.  
The stable and unstable equilibria of \eqref{eq:Rtilde_dynamics} are
\begin{align*}
    \mathcal{E}_s &= \{ I_3 \}, &
    \mathcal{E}_u &= \{U \in \SO(3) \mid \mathrm{tr}(U) = -1\}.
\end{align*}
\end{lemma}

\begin{remark}
It is well known that almost–global asymptotic stability is the strongest form of stability that can be achieved by a smooth attitude observer due to the topological obstruction on $\SO(3)$ \citep{bhat2000topological}.
\end{remark}

Condition \eqref{eq:PE eta} is equivalent to requiring that the time-derivative $\dot{\bm{\eta}}_{\mathcal{I}}$ is uniformly persistently exciting, i.e., there exist constants $\delta, \epsilon > 0$ such that, for all $t \geq 0$, $\|\dot{\bm{\eta}}_{\mathcal{I}}(s)\| > \epsilon$ for some $s \in [t, t+ \delta]$. Since $\bm{\eta}_{\mathcal{I}}$ evolves according to $ \dot{\bm{\eta}}_{\mathcal{I}} = \dot{Q}_T \bm{e}_3 = Q_T \bm{\omega}_T^\times \bm{e}_3$, a necessary condition for \eqref{eq:PE eta} is that there must exist $\delta >0$ and $\epsilon>0$ such that for all $t \geq 0$,
$$\|\bm{\omega}_T(s) \times \bm{e}_3 \| > \epsilon, \quad s \in [t, t+ \delta]. $$
The case where this is not satisfied is considered in Subsection \ref{sec:coupled_observer}.

\subsection{Relative position and velocity observer}

Let $\hat{\bm{\xi}}, \hat{\bm{v}} \in \mathbb{R}^3$ denote the estimates of the relative position $\bm{\xi}$ and velocity $\bm{v}$, respectively. From the dynamics in \eqref{eq:dyn_p}-\eqref{eq:dyn_v}, the observer is defined as
\begin{equation} \label{pv_observer}
    \begin{cases}
        \dot{\hat{\bm{\xi}}} = -\bm{\omega}_B^\times  \hat{\bm{\xi}} + \hat{\bm{v}} + \bm{\sigma}_\xi\\
        \dot{\hat{\bm{v}}} = -\bm{\omega}_B^\times  \hat{\bm{v}} + \bm{a}_B - \hat{R}^\top \bm{a}_T + \bm{\sigma}_v,
    \end{cases}
\end{equation}
with innovation terms $\bm{\sigma}_{\xi},\bm{\sigma}_{v}\in \mathbb{R}^3$ to be determined.

Define the errors $\tilde{\bm{\xi}} = \bm{\xi} - \hat{\bm{\xi}} \in \R^3$ and $\tilde{\bm{v}} = \bm{v} - \hat{\bm{v}} \in \R^3$. Their dynamics then follow as
\begin{align} \label{eq:p_error}
        \dot{\tilde{\bm{\xi}}} &= -\bm{\omega}_B^\times  \tilde{\bm{\xi}} + \tilde{\bm{v}} - \bm{\sigma}_\xi. \\
    \dot{\tilde{\bm{v}}} &= -\bm{\omega}_B^\times  \tilde{\bm{v}} - R^\top \bm{a}_T +\hat{R}^\top \bm{a}_T - \bm{\sigma}_v \notag \\ \label{eq:v_error}
    &= -\bm{\omega}_B^\times  \tilde{\bm{v}} - \bm{\sigma}_v + \hat{R}^\top (I_3 - \tilde{R}) \bm{a}_T.
\end{align}

These dynamics show that the residual term $\hat{R}^\top (I_3 - \tilde{R}) \bm{a}_T$ represents the effect of the unknown rotation error on the velocity dynamics that vanishes as $\tilde{R} \rightarrow I_3$.
Introducing the error $ \bm{x} = (\tilde{\bm{\xi}}, \tilde{\bm{v}}) \in \R^6$,
\eqref{eq:p_error}-\eqref{eq:v_error} can be written as
\begin{align}\label{pv error}
    \dot{\bm{x}} = A(t) \bm{x} - \begin{bmatrix}
        \bm{\sigma}_\xi \\ \bm{\sigma}_v
    \end{bmatrix} + \bm{r}(\tilde{R}, \hat{R}, \bm{a}_T), 
\end{align}
where
$$ A(t) = \begin{bmatrix}
    -\bm{\omega}_B^\times  & I_3 \\ 0_{3,3} & - \bm{\omega}_B^\times 
\end{bmatrix}, \quad \bm{r}(\tilde{R}, \hat{R}, \bm{a}_T) = \begin{bmatrix}
    0_{3,1} \\ \hat{R}^\top (I_3 - \tilde{R}) \bm{a}_T
\end{bmatrix}. $$
The available measurement is the bearing \eqref{eq:bearing}. Using the identity $\pi_{\bm{y}_\xi} \bm{\xi} = 0$, it follows that $\pi_{\bm{y}_\xi} ( \hat{\bm{\xi}} + \tilde{\bm{\xi}}) = 0$, which yields the output equation
\begin{align} \label{pv output}
    \bm{y} &= C(t) \bm{x},
\end{align}
with $\bm{y}(t) = - \pi_{\bm{y}_\xi} \hat{\bm{\xi}}$, and $C(t) = \begin{bmatrix}
    \pi_{\bm{y}_\xi} & 0_{3,3}
\end{bmatrix}$. 

Therefore, equations \eqref{pv error}-\eqref{pv output} define an LTV system forced by the external perturbation $\bm{r}(\tilde{R}, \hat{R}, \bm{a}_T)$ induced by the attitude error $\tilde{R}$.

We first consider the unforced system obtained by setting 
$\bm{r}(\tilde{R}, \hat{R}, \bm{a}_T) \equiv 0$, corresponding to the case where the attitude observer \eqref{R_observer} has converged, as guaranteed by Lemma~\ref{lemma_attitude_observer}. Then, the innovation terms can be designed to stabilize the error dynamics and drive $\bm{x} \rightarrow 0$ using a Riccati-based observer gain as follows
    \begin{align} \label{sigma_pv}
             \bm{\sigma}_{\xi} &= K_{\xi} \bm{y}, &
             \bm{\sigma}_{v} &= K_{v} \bm{y},
    \end{align}
    with the gain matrix given by $K = [
        K_{\xi}^\top, K_{v}^\top ]^\top = PC^\top D$, where $P$ is the solution to the continuous Riccati equation
    \begin{equation} \label{eq:cre}
        \dot{P} = AP + PA^\top - PC^\top D CP + S, \quad P(0) \in \mathbb{S}_+(6),
    \end{equation}
    with $S \in \mathbb{S}_+(6)$ and $D \in \mathbb{S}_+(3)$ bounded and continuous symmetric positive definite matrix-valued functions.
    The resulting closed-loop error dynamics are
    \begin{equation}
        \dot{\bm{x}} = (A(t) - K(t) C(t)) \bm{x}.
    \end{equation}  
Provided that the pair $(A(t),C(t))$ is uniformly observable, and under the positive definiteness assumptions on 
$S$ and $D$, the Riccati equation \eqref{eq:cre} admits a unique solution $P(t)$ 
that remains bounded, uniformly positive definite, and well-conditioned for all $t \geq 0$ (cf.~\cite[Lemma 2.5, Lemma 2.6]{hamel2017position}). This, in turn, guarantees global exponential stability of the equilibrium $\bm{x} = 0$.

The following lemma (see Appendix for proof) establishes a persistently exciting condition that guarantees uniform observability of the pair $(A(t),C(t))$. 
    \begin{lemma} \label{lemma_uniform_obs}
  Assume that the angular velocities $\bm{\omega}_B$ and $\bm{\omega}_T$ are uniformly continuous and bounded, and that the bearing vector in the inertial-frame $\bm{y}_\xi^{\mathcal{I}} := Q_B \bm{y}_\xi \in \mathrm{S}^2$ is persistently exciting in the sense that there exist $\delta, \mu > 0$ such that, for all $t \geq 0$,
    \begin{equation} \label{eq:PE pv}
        \frac{1}{\delta} \int_t^{t+\delta} \pi_{\bm{y}_\xi^{\mathcal{I}}(s)} d s \succeq \mu I_3.
     \end{equation}
     Then, the pair $(A(t), C(t))$ is uniformly observable, and $\bm{x} = 0$ is globally exponentially stable. 
    \end{lemma}

 \subsection{Cascade observer}

We now combine the results in Lemmas~\ref{lemma_attitude_observer} and \ref{lemma_uniform_obs} to examine the interconnection. 
    \begin{theorem} \label{thm:main_theorem}
        Consider the attitude observer \eqref{R_observer} with innovation \eqref{sigma_R} and the Riccati observer \eqref{pv_observer} with innovations \eqref{sigma_pv}.
        Assume the angular velocities $\bm{\omega}_B$ and $\bm{\omega}_T$ are uniformly continuous and bounded, and the target acceleration $\bm{a}_T$ is bounded. Under the persistently exciting conditions \eqref{eq:PE eta} and \eqref{eq:PE pv}, the equilibrium $(\tilde{R}, \bm{x}) = (I_3, 0)$ of the interconnection system
        \begin{equation}\label{eq:cascade error system}
            \begin{aligned}
                \dot{\tilde{R}} &= [\tilde{R}, \bm{\omega}_T^\times ]  + 2k_R ((\hat{R} \bm{\eta}) \times \bm{e}_3)^\times \tilde{R} \\
                \dot{\bm{x}} &= (A(t) - K(t) C(t)) \bm{x} + \bm{r}(\tilde{R}, \hat{R}, \bm{a}_T)
            \end{aligned}
        \end{equation}
        is almost globally asymptotically stable and locally exponentially stable.
    \end{theorem}
    \begin{proof}
        Condition \eqref{eq:PE pv} in Lemma~\ref{lemma_uniform_obs} guarantees that the equilibrium $(\tilde{\bm{\xi}}, \tilde{\bm{v}}) = (0,0)$ of the unforced subsystem \eqref{pv error} (when the external perturbation satisfies $\bm{r}(\tilde{R}, \hat{R}, \bm{a}_T) \equiv 0$) is uniformly globally exponentially stable. 
        Moreover, since the acceleration $\bm{a}_T(t)$ is bounded and both $\tilde{R}$ and $\hat{R}$ remain bounded on the compact set $\mathbf{SO}(3)$, the forced system \eqref{pv error} is input-to-state stable (ISS) w.r.t. the equilibrium $(\tilde{\bm{\xi}}, \tilde{\bm{v}}) = (0,0)$ and the input $\bm{r}(\tilde{R}, \hat{R}, \bm{a}_T)$ (see \cite[Lemma 4.6]{Khalil2002}). 
        
        Since the $\bm{x}$-dynamics are globally ISS w.r.t. the equilibrium $(\tilde{\bm{\xi}}, \tilde{\bm{v}}) = (0,0)$ and the input $\bm{r}(\tilde{R}, \hat{R}, \bm{a}_T)$, and the $\tilde{R}$-dynamics are almost globally asymptotically stable w.r.t. the equilibrium $\tilde{R} = I_3$, the almost global asymptotic stability claim of the overall interconnection follows directly.

        Local exponential stability follows by invoking standard results. 
        The unforced $\bm{x}$-dynamics, with $\bm{r}(\tilde{R},\hat{R},\bm{a}_T)\equiv 0$, are locally exponentially stable and thus locally exponentially ISS with respect to $\bm{r}(\tilde{R},\hat{R},\bm{a}_T)$. 
        The cascade interconnection of a locally exponentially stable system with a locally exponentially ISS system is locally exponentially stable. 
    \end{proof}

\subsection{Observer design with coupled dynamics} \label{sec:coupled_observer}
In the previous section, the stability of the Riccati observer relied on the assumption that the inertial normal direction $\bm{\eta}_{\mathcal{I}}$ satisfies the persistently exciting condition \eqref{eq:PE eta}. In many practical landing scenarios, however, $\bm{\eta}_{\mathcal{I}}$ may remain constant or slowly time-varying. In such cases, the attitude and translational dynamics cannot be fully decoupled.

If the persistence of excitation condition given in Lemma \ref{lemma_attitude_observer} is not satisfied then the proposed complementary filter ensures only that $\tilde{R}\bm{e}_3 \rightarrow \bm{e}_3$. 
It is possible to overcome this challenge by constructing an innovation along the normal direction $\bm{e}_3$ based on condition \eqref{eq:secondPEcondition}. 

Recall the attitude observer \eqref{R_observer} with a new scalar innovation $\sigma_\theta$ in direction $\bm{e}_3$:
\begin{equation} \label{eq:attitude_obs_theta}
    \dot{\hat{R}} =  - \bm{\omega}_T^\times \hat{R} + \hat{R} \bm{\omega}_B^\times + (\bm{\sigma}_R - \sigma_{\theta} \bm{e}_3)^\times \hat{R}. 
\end{equation}
Let us rewrite the velocity error dynamics as follows: 
\begin{align}
    \dot{\tilde{\bm{v}}} &= -\bm{\omega}_B^\times  \tilde{\bm{v}} - \bm{\sigma}_v + \hat{R}^\top (I_3 - \tilde{R}) \bm{a}_T \notag \\
    &= -\bm{\omega}_B^\times  \tilde{\bm{v}} - \bm{\sigma}_v + \hat{R}^\top (I_3 - \tilde{R}) (\bm{e}_3 \bm{e_3}^\top + \pi_{\bm{e}_3}) \bm{a}_T \notag \\
    &= -\bm{\omega}_B^\times  \tilde{\bm{v}} - \bm{\sigma}_v + \hat{R}^\top (I_3 - \tilde{R}) \pi_{\bm{e}_3} \bm{a}_T + \bm{r}_{\bm{e}_3}(\tilde{R}, \hat{R}, \bm{a}_T). \notag 
\end{align}
where $\bm{r}_{\bm{e}_3}(\tilde{R}, \hat{R}, \bm{a}_T) = \hat{R}^\top  (\bm{e_3} - \tilde{R} \bm{e}_3)\bm{e_3}^\top \bm{a}_T$ goes to zero as $\tilde{R}\bm{e}_3 \rightarrow \bm{e}_3$ and can be treated as a vanishing perturbation. 

The component $\sigma_\theta \bm{e}_3$ is orthogonal to $\bm{\sigma}_R$ by construction and therefore does not affect the convergence of $\tilde{R}\bm{e}_3$ to $\bm{e}_3$. 
From there, $\bm{\sigma}_R$ as defined in \eqref{sigma_R} ensures that $\tilde{R}$ converges to a rotation  
$\tilde{R}^\star = \exp(\tilde{\theta} \bm{e}_3^\times) \in \SO(3)$ for some constant angle $\tilde{\theta} \in \mathrm{S}^1$ around the $\bm{e}_3$ axis. 
Assuming that $\bm{r}_{\bm{e}_3} \equiv 0$ (i.e, $\tilde{R} \rightarrow \tilde{R}^\star$), one can write 
\begin{align*}
    \dot{\tilde{\bm{v}}} 
    &= -\bm{\omega}_B^\times  \tilde{\bm{v}} - \bm{\sigma}_v + \hat{R}^\top (I_3 - \exp(\tilde{\theta} \bm{e}_3^\times)) \bm{a}_T. 
    \notag 
\end{align*}
Using Rodrigues' formula \eqref{eq:rodrigues_formula}, we obtain the first-order approximation $\exp(\tilde{\theta} \bm{e}_3^\times) = I_3 + \tilde{\theta} \bm{e}_3^\times + O(|\tilde{\theta}|^2)$. Then, noting that $\bm{e}_3^\times \pi_{\bm{e}_3} = \bm{e}_3^\times$, the velocity error dynamics can be approximated as
    \begin{align}
    \dot{\tilde{\bm{v}}} &= -\bm{\omega}_B^\times  \tilde{\bm{v}} - \bm{\sigma}_v - \hat{R}^\top (\tilde{\theta} \bm{e}_3^\times) \bm{a}_T + O(|\theta|^2) \notag \\
    &= -\bm{\omega}_B^\times  \tilde{\bm{v}} - \bm{\sigma}_v + \hat{R}^\top \bm{a}_T^\times \bm{e}_3  \tilde{\theta} + O(|\tilde{\theta}|^2) 
\end{align}

Defining $ \bm{x}_\theta = (\tilde{\bm{\xi}}, \tilde{\bm{v}}, \tilde{\theta}) \in \R^6 \times \R$, and $\bm{\sigma} = (\bm{\sigma}_\xi, \bm{\sigma}_v, \sigma_\theta)$,  
the full error dynamics can be written as
\begin{align}\label{pv error2}
    \dot{\bm{x}}_\theta &= A_\theta(\bm{x}, t) \bm{x}_\theta - \bm{\sigma}
    + O(|\theta|^2) &
    \bm{y} &= C_\theta(t) \bm{x}_\theta, 
\end{align}
where
$$ A_\theta(t) = \begin{bmatrix}
    -\bm{\omega}_B^\times  & I_3 & 0_{3,1} \\ 0_{3,3} & - \bm{\omega}_B^\times & \hat{R}^\top \bm{a}_T^\times \bm{e}_3 \\ 0_{1,3} & 0_{1,3} & 0
\end{bmatrix}, \;\; C_\theta(t) = \begin{bmatrix}
    \pi_{\bm{y}_\xi} & 0_{3,4}
\end{bmatrix}, $$
with innovations chosen as 
\begin{align*} 
             \bm{\sigma}_{\xi} &= K_{\xi} \bm{y}, &
             \bm{\sigma}_{v} &= K_{v} \bm{y}, & \sigma_{\theta} = K_{\theta} \bm{y}.
\end{align*}
where $K = [ K_{\xi}^\top, K_{v}^\top, K_\theta^\top ]^\top = P C_\theta^\top D$, with $P \in \mathbb{S}_+(7)$ solution to the continuous Riccati equation \eqref{eq:cre} associated with the matrices $A_\theta$ and $C_\theta$. 

\begin{assumption} \label{assump1}
Assume that $\bm{a}_T$ is continuous and bounded and define $\bm{b}(t) := \bm{a}_T^{\mathcal{I}} \times \bm{\eta}_{\mathcal{I}}$, with $\bm{a}_T^{\mathcal{I}} = Q_T \bm{a}_T$ the target's specific acceleration expressed in the inertial frame. 
Suppose that $\bm{b}(t)$ admits the decomposition $\bm{b}(t) = \bm{b}_0 + \bm{b}_1(t)$, such that $\bm{b}_0 \in \R^3$ is a (non-zero) constant vector, and $\bm{b}_1(t)$ is the time-varying component of $\bm{b}$ that satisfies for constants $\delta, \mu > 0$ defined in \eqref{eq:PE pv} and for all $t \geq 0$:
\begin{align}
    \frac{1}{\delta} \int_{t}^{t+\delta} \| \pi_{\bm{y}_{\xi}^{\mathcal{I}}} 
    \bm{b}_1(t) \|^2ds \leq \frac{\mu}{4} \|\bm{b}_0\|^2.
    \label{eq:secondPEcondition}
\end{align}

\end{assumption}

The next lemma (see Appendix for proof) provides sufficient conditions to ensure uniform observability of $(A_\theta, C_\theta)$, which in turn guarantees the local exponential stability of system~\eqref{pv error2} at $\bm{x}_\theta = 0$. 

\begin{lemma} \label{lemma_pv_yaw_uo}
Consider system \eqref{pv error2} and assume that the angular velocities $\bm{\omega}_B$ and $\bm{\omega}_T$ are uniformly continuous and bounded. Suppose that there exist constants $\delta, \mu >0$ such that condition \eqref{eq:PE pv} in Lemma~\ref{lemma_uniform_obs} holds, 
and assume further that Assumption~\ref{assump1} holds.
Then, the pair $(A_\theta, C_\theta)$ is uniformly observable and the equilibrium $(\tilde{\bm{\xi}}, \tilde{\bm{v}}, \tilde{\theta}) = 0$ is locally exponentially stable.    
\end{lemma}

The cascade formed by subsystems $\tilde{R}\bm{e}_3$ and $\bm{x}_\theta$ can be examined similarly to the analysis in Theorem~\ref{thm:main_theorem}. Since the $\tilde{R}\bm{e}_3$-dynamics are locally exponentially stable, and the unforced subsystem~\eqref{pv error2} is locally exponentially stable (and therefore locally exponentially ISS with respect to the bounded vanishing perturbation $\bm{r}_{e_3}$), the equilibrium of the overall cascade is locally exponentially stable.

\section{Simulation Results} \label{sec:simulations}

In this section, we provide simulation results to evaluate the performance of the observers presented in Section \ref{sec:cascade_observer}.

We simulate the cascade observer \eqref{R_observer}--\eqref{pv_observer} under the following motion scenario: The UAV follows a circular horizontal path with measured angular rate and specific acceleration set to $\bm{\omega}_B =  [0, 0, 0.5]^\top$, 
$\bm{a}_B = \bm{\omega}_B^\times \bm{v}_B - g Q_B^\top \bm{e}_3$. The target undergoes an oscillatory (roll) motion
with angular velocity $\bm{\omega}_T = [-1.5 \sin(t), 0, 0]^\top$, which ensures the \emph{p.e.} condition \eqref{eq:PE eta} is satisfied,
and has no translational motion, i.e., the specific acceleration is
$\bm{a}_T = -g Q_T^\top \bm{e}_3$.
The attitude observer gain is set to $k_R =2.5$. For the Riccati observer \eqref{pv_observer}: $P(0)= 2 I_6$, $D=50I_3$ and $S(t) = \mathrm{diag}(0.005 I_3, 0.01 \gamma I_3)$, with $\gamma = \|\bm{a}_T\|^2 (1-\hat{\eta}^\top\eta + 10^{-2})$.
This choice of $S$ allows us to avoid relying on the model initially when $\hat{R}$ is inaccurate to reduce the effect of large perturbations, and increases it as the attitude converges.

For the observer with coupled dynamics proposed in subsection \ref{sec:coupled_observer}, the target's motion is defined as follows: $\bm{\omega}_T = [0, 0, -0.8]^\top$ and  $\bm{a}_T = \bm{\omega}_T^\times \bm{v}_T -g Q_T^\top \bm{e}_3$.
The Riccati initial state and gains are: $P(0)=  \mathrm{diag}(2 I_6, 0.1)$, $D=30I_3$ and $S(t) = \mathrm{diag}(0.005 I_3, 0.005 \gamma I_3, 0.001)$.
 
We perform 50 Monte Carlo runs for each scenario. Initial estimation errors are randomly generated from Gaussian distributions:
the initial relative attitude $\hat{R}(0)$ corresponds to an average error of $45^\circ$ per rotation axis with a $30^\circ$ standard deviation,  
the average initial relative position is $\hat{\bm{\xi}}(0) = [-4.5, -5, 6]^\top$ with standard deviation $5$ per axis,  and the average initial relative velocity is $\hat{\bm{v}}(0) = [2, -1.5, 0.5]^\top$ with standard deviation $2$ per axis.

\subsection{Results and discussion}
\begin{figure}[ht]
    \centering
\includegraphics[width=.8\linewidth,trim= 50 0 40 0,clip]{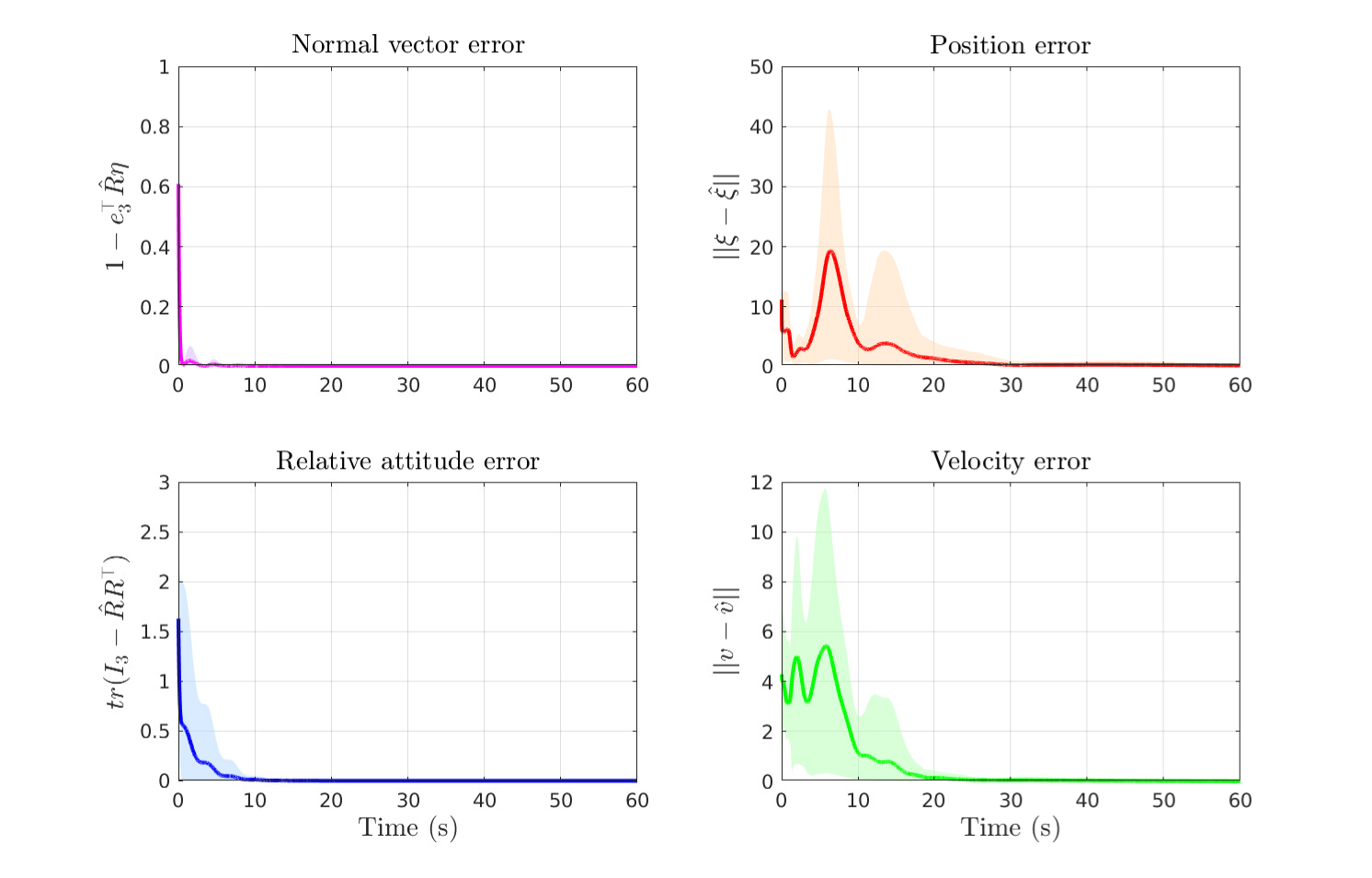}
    \caption{Monte Carlo results for the cascade observer \eqref{R_observer}-\eqref{pv_observer}. (left) reduced attitude and attitude estimation errors. (right) position and velocity errors.
    Solid lines show the median over all initial conditions; shaded regions indicate the $5$th--$95$th percentiles.}
    \label{fig:cascade_errors}
\end{figure}

Fig.~\ref{fig:cascade_errors} illustrates the normal direction and orientation estimation errors computed respectively as $1 - \bm{e}_3^\top \hat{R}\bm{\eta}$, and
$\mathrm{tr}(I_3 - \hat{R} R^\top)$ on the left side. The right-hand plots display the errors on the relative position $\|\bm{\xi} - \hat{\bm{\xi}}\|$ and velocity $\|\bm{v} - \hat{\bm{v}}\|$.
The shaded areas illustrate the $5$th-$95$th percentile of error across all initial conditions.
The plots show
that the normal and attitude errors converge to zero. 
A short transient increase is visible in the position and velocity errors, which is expected before $\hat{R}$ has converged to $R$. Once the attitude error decays, the perturbation vanishes and the position-velocity estimates rapidly converge.
Overall, all error trajectories converge to zero, confirming the asymptotic stability of the proposed observer.

\begin{figure}[ht]
    \centering
\includegraphics[width=.8\linewidth,trim= 50 0 40 0,clip]{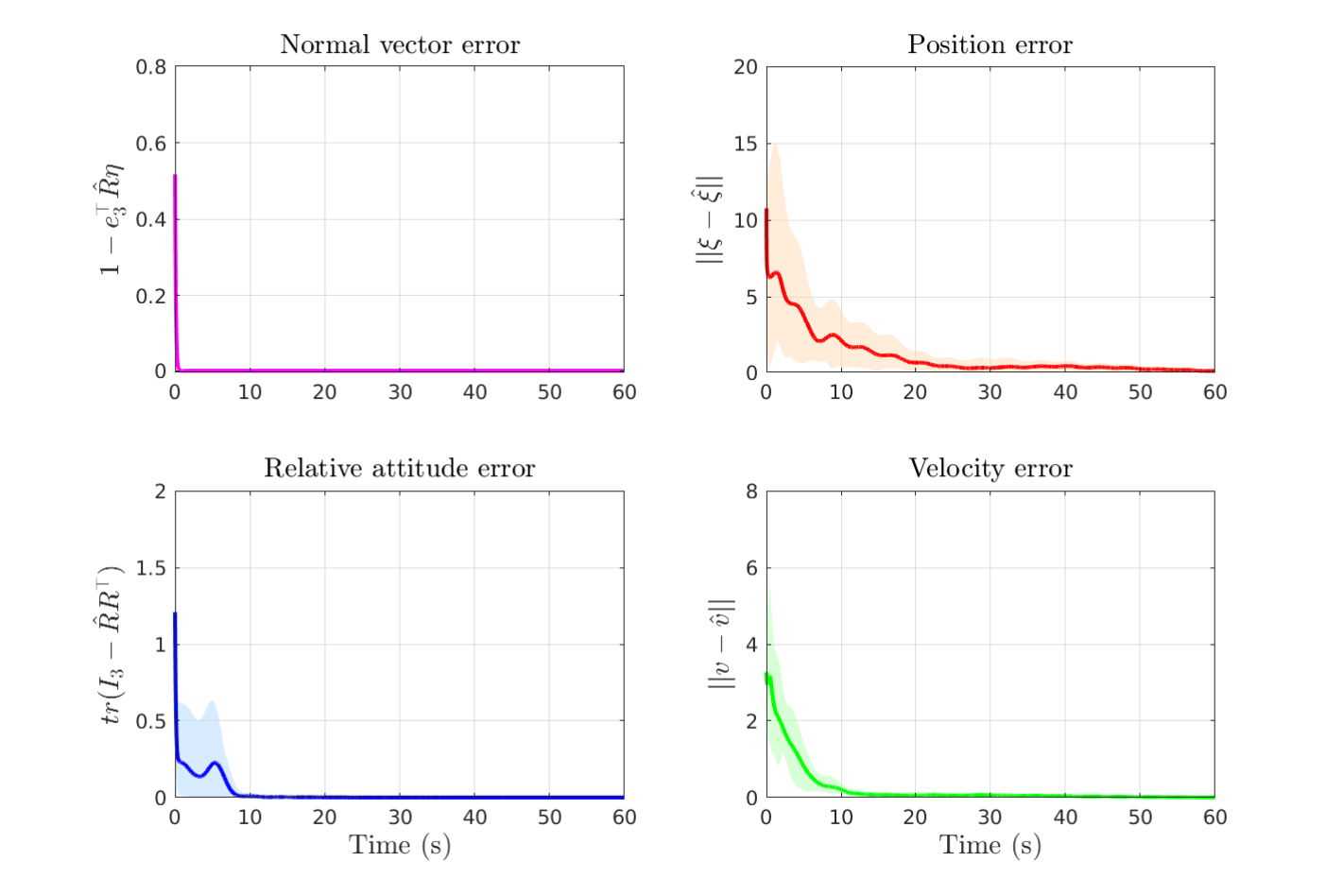}
    \caption{Monte Carlo results for the coupled observer \eqref{eq:attitude_obs_theta}. (left) reduced attitude and attitude estimation errors. (right) position and velocity errors.}
    \label{fig:coupled_errors}
\end{figure}

The estimation errors for the observer with coupled position-velocity and yaw dynamics are illustrated in Fig.~\ref{fig:coupled_errors}. It can be seen that although the normal $\bm{\eta}_{\mathcal{I}}$ does not satisfy the \emph{p.e.} condition \eqref{eq:PE eta}, the full attitude estimate $\hat{R}$ converges to $R$ (as $\tilde{\theta}$ converges to zero). Additionally, the position and velocity estimates converge to the true states, which demonstrates the observer's performance.


\section{Conclusion} \label{sec:conclusion}

The problem of estimating the relative pose and velocity between a UAV and a moving planar platform was addressed using dual-IMU and monocular vision sensing. A cascade observer was proposed, combining a relative attitude observer on $\SO(3)$ and a Riccati-based estimator for position and velocity. The approach was shown to achieve almost global asymptotic and locally exponential convergence under persistent excitation, and an additional analysis demonstrated that, when the platform rotates only about its surface normal, its measured linear acceleration can be used to recover the remaining unobservable angle. The effectiveness of the proposed architecture was validated through extensive simulations.

Future work will focus on integrating the observer into a control framework for autonomous approach and landing, and on validating its performance on UAV-landing pad experiments under realistic conditions.
Further efforts will focus on improving robustness to communication delays and to IMU imperfections, including sensor biases and synchronization errors.

\bibliographystyle{unsrtnat}
\bibliography{ref}

@inproceedings{falanga2017vision,
  title={Vision-based autonomous quadrotor landing on a moving platform},
  author={Falanga, Davide and Zanchettin, Alessio and Simovic, Alessandro and Delmerico, Jeffrey and Scaramuzza, Davide},
  booktitle={2017 IEEE International Symposium on Safety, Security and Rescue Robotics (SSRR)},
  pages={200--207},
  year={2017},
  organization={IEEE}
}

@inproceedings{lee2012autonomous,
  title={Autonomous landing of a VTOL UAV on a moving platform using image-based visual servoing},
  author={Lee, Daewon and Ryan, Tyler and Kim, H Jin},
  booktitle={2012 IEEE international conference on robotics and automation},
  pages={971--976},
  year={2012},
  organization={IEEE}
}

@article{serra2016landing,
  title={Landing of a quadrotor on a moving target using dynamic image-based visual servo control},
  author={Serra, Pedro and Cunha, Rita and Hamel, Tarek and Cabecinhas, David and Silvestre, Carlos},
  journal={IEEE Transactions on Robotics},
  volume={32},
  number={6},
  pages={1524--1535},
  year={2016},
  publisher={IEEE}
}

@article{wenzel2011automatic,
  title={Automatic take off, tracking and landing of a miniature UAV on a moving carrier vehicle},
  author={Wenzel, Karl Engelbert and Masselli, Andreas and Zell, Andreas},
  journal={Journal of intelligent \& robotic systems},
  volume={61},
  number={1},
  pages={221--238},
  year={2011},
  publisher={Springer}
}

@article{borowczyk2017autonomous,
  title={Autonomous landing of a multirotor micro air vehicle on a high velocity ground vehicle},
  author={Borowczyk, Alexandre and Nguyen, Duc-Tien and Phu-Van Nguyen, Andr{\'e} and Nguyen, Dang Quang and Saussi{\'e}, David and Le Ny, Jerome},
  journal={Ifac-Papersonline},
  volume={50},
  number={1},
  pages={10488--10494},
  year={2017},
  publisher={Elsevier}
}

@inproceedings{mellinger2010control,
  title={Control of quadrotors for robust perching and landing},
  author={Mellinger, Daniel and Shomin, Michael and Kumar, Vijay},
  booktitle={Proceedings of the international powered lift conference},
  pages={205--225},
  year={2010}
}

@article{kang2018precision,
  title={A precision landing test on motion platform and shipboard of a tilt-rotor UAV based on RTK-GNSS},
  author={Kang, Youngshin and Park, Bum-Jin and Cho, Am and Yoo, Chang-Sun and Kim, Yushin and Choi, Seongwook and Koo, Sam-Ok and Oh, Soohun},
  journal={International Journal of Aeronautical and Space Sciences},
  volume={19},
  number={4},
  pages={994--1005},
  year={2018},
  publisher={Springer}
}

@inproceedings{acuna2018vision,
  title={Vision-based UAV landing on a moving platform in GPS denied environments using motion prediction},
  author={Acuna, Raul and Zhang, Ding and Willert, Volker},
  booktitle={2018 Latin American Robotic Symposium, 2018 Brazilian Symposium on Robotics (SBR) and 2018 Workshop on Robotics in Education (WRE)},
  pages={515--521},
  year={2018},
  organization={IEEE}
}

@article{cui2023coarse,
  title={Coarse-to-fine visual autonomous unmanned aerial vehicle landing on a moving platform},
  author={Cui, Qiangqiang and Liu, Min and Huang, Xiaoyin and Gao, Ming},
  journal={Biomimetic Intelligence and Robotics},
  volume={3},
  number={1},
  pages={100088},
  year={2023},
  publisher={Elsevier}
}

@article{lepetit2009ep,
  title={EP n P: An accurate O (n) solution to the P n P problem},
  author={Lepetit, Vincent and Moreno-Noguer, Francesc and Fua, Pascal},
  journal={International journal of computer vision},
  volume={81},
  number={2},
  pages={155--166},
  year={2009},
  publisher={Springer}
}

@inproceedings{bouazza2025simple,
  title={A Simple Algebraic Solution for Estimating the Pose of a Camera from Planar Point Features},
  author={Bouazza, Tarek and Hamel, Tarek and Samson, Claude},
  booktitle={2025 IEEE/RSJ International Conference on Intelligent Robots and Systems (IROS)},
  pages={20189--20194},
  year={2025},
  organization={IEEE}
}

@article{hamel2017position,
  title={Position estimation from direction or range measurements},
  author={Hamel, Tarek and Samson, Claude},
  journal={Automatica},
  volume={82},
  pages={137--144},
  year={2017},
  publisher={Elsevier}
}

@article{trumpf2012analysis,
  title={Analysis of non-linear attitude observers for time-varying reference measurements},
  author={Trumpf, Jochen and Mahony, Robert and Hamel, Tarek and Lageman, Christian},
  journal={IEEE Transactions on Automatic Control},
  volume={57},
  number={11},
  pages={2789--2800},
  year={2012},
  publisher={IEEE}
}

@book{Khalil2002,
   author = {H. K. Khalil},
   edition = {3rd},
   isbn = {9780130673893},
   publisher = {Prentice Hall},
   title = {Nonlinear Systems},
   year = {2002}
}

@article{bhat2000topological,
  title={A topological obstruction to continuous global stabilization of rotational motion and the unwinding phenomenon},
  author={Bhat, Sanjay P and Bernstein, Dennis S},
  journal={Systems \& control letters},
  volume={39},
  number={1},
  pages={63--70},
  year={2000},
  publisher={Elsevier}
}

@book{chen1984linear,
   author = {Chi-Tsong. Chen},
   publisher = {Saunders College Publishing},
   title = {Linear system theory and design},
   year = {1984}
}

@article{hamel2017riccati,
  title={Riccati observers for the nonstationary PnP problem},
  author={Hamel, Tarek and Samson, Claude},
  journal={IEEE Transactions on Automatic Control},
  volume={63},
  number={3},
  pages={726--741},
  year={2017},
  publisher={IEEE}
}

@inproceedings{morin2017uniform,
  title={Uniform observability of linear time-varying systems and application to robotics problems},
  author={Morin, Pascal and Eudes, Alexandre and Scandaroli, Glauco},
  booktitle={International conference on geometric science of information},
  pages={336--344},
  year={2017},
  organization={Springer}
}

\appendix

\section{Proofs}
\begin{proof}[Proof of Lemma \ref{lemma_attitude_observer}]
To show that $\tilde{R}$ converges to $I_3$ almost globally asymptotically and locally exponentially under the persistently exciting condition \eqref{eq:PE eta}, we exploit the closure of $\SO(3)$ under conjugation and define the following transformed attitude error
$$\tilde{R}_{Q_T} := Q_T \hat{R} R^\top Q_T^\top = Q_T  \tilde{R} Q_T^\top  \in \SO(3),$$ 
such that convergence of $\tilde{R}_{Q_T}$ to $I_3$ is equivalent to the convergence of $\tilde{R}$ to $I_3$, because conjugation by $Q_T$ is an isometry of $\SO(3)$.  
Using \eqref{pose kinematics}-\eqref{eq:Rtilde_dynamics}, the dynamics of $\tilde{R}_{Q_T}$ are 
\begin{align} \label{erro R dyn}
        \dot{\Tilde{R}}_{Q_T} &=  Q_T \left([\tilde{R}, \bm{\omega}_T^\times ]  + \bm{\sigma}_R^\times  \tilde{R} \right) Q_T^\top + Q_T (\bm{\omega}_T^\times \tilde{R} - \tilde{R} \bm{\omega}_T^\top ) Q_T^\top  \notag \\
        &= Q_T [\tilde{R}, \bm{\omega}_T^\times ] Q_T^\top - Q_T [\tilde{R}, \bm{\omega}_T^\times ] Q_T^\top  + Q_T \bm{\sigma}_R^\times \tilde{R}  Q_T^\top \notag \\
        &=  Q_T \bm{\sigma}_R^\times Q_T^\top Q_T \tilde{R} Q_T^\top \notag \\
        &= \bar{\bm{\sigma}}_R^\times \tilde{R}_{Q_T}.
\end{align}
where $\bar{\bm{\sigma}}_R := Q_T \bm{\sigma}_R$, which can be further expressed as 
\begin{align*}
    \bar{\bm{\sigma}}_{R} = 2k_R Q_T ((\hat{R}\bm{\eta}) \times \bm{e}_3) &= 2k_R (Q_T (\hat{R}\bm{\eta}) \times Q_T^\top Q_T \bm{e}_3) \\ 
    &= 2k_R ( \hat{\bm{\eta}}_{\mathcal{I}} \times \bm{\eta}_{\mathcal{I}}).
\end{align*}
with $\hat{\bm{\eta}}_{\mathcal{I}} := Q_T \hat{R}\bm{\eta}$ denoting the estimate of the normal to the target expressed in the inertial frame.
Then, this is a classical problem of ensuring $\Tilde{R}_{Q_T} \rightarrow I_3$ given the time-variyng reference direction measurement $\bm{\eta}_{\mathcal{I}}$ \citep{trumpf2012analysis}.
Consider the Lyapunov function $\bar{V}_R(\tilde{R}_{Q_T}) := \mathrm{tr}(I_3 - \tilde{R}_{Q_T})$, which is smooth on $\SO(3)$ and satisfies $\bar{V}_R(\tilde{R}_{Q_T})=0$ if and only if $\tilde R_{Q_T}=I_3$. computing its time derivative yields 
\begin{align*}
    \dot{\bar{V}}_R &= - \mathrm{tr}(\dot{\tilde{R}}_{Q_T}) = - 2k_R \mathrm{tr}\left( ( \hat{\bm{\eta}}_{\mathcal{I}} \times \bm{\eta}_{\mathcal{I}})^\times \tilde{R}_{Q_T}\right). 
\end{align*}  
Using the identity $(a \times b)^\times = ba^\top - ab^\top$, we obtain
\begin{align}
\dot{V}_R &= -2k_R \mathrm{tr}\left(( \bm{\eta}_{\mathcal{I}} \hat{\bm{\eta}}_{\mathcal{I}}^\top - \hat{\bm{\eta}}_{\mathcal{I}} \bm{\eta}_{\mathcal{I}}^\top) \tilde{R}_{Q_T}\right), \notag \\
&= -2k_R \mathrm{tr}(\bm{\eta}_{\mathcal{I}} \bm{\eta}_{\mathcal{I}}^\top -  \tilde{R}_{Q_T} \bm{\eta}_{\mathcal{I}} \bm{\eta}_{\mathcal{I}}^\top \tilde{R}_{Q_T} ), \notag \\
&= -2k_R \bm{\eta}_{\mathcal{I}}^\top (I_3 -  \tilde{R}_{Q_T}^2) \bm{\eta}_{\mathcal{I}} ,  \notag\\
&= -k_R \bm{\eta}_{\mathcal{I}} ^\top (I_3 -  \tilde{R}_{Q_T}^2)^\top (I_3 -  \tilde{R}_{Q_T}^2) \bm{\eta}_{\mathcal{I}}, \notag \\
&= -k_R \|(I_3 -  \tilde{R}_{Q_T}^2) \bm{\eta}_{\mathcal{I}} \|^2 \leq 0.
\end{align}
This shows that $\bar{V}_R$ is non-increasing and that $\dot{\bar{V}}_R=0$ only when $(I_3 - \tilde{R}_{Q_T}^2)\eta_{\mathcal I}=0$.
The remainder of the proof is a direct application of \citep[Proposition 4.6]{trumpf2012analysis}.
Under the persistently exciting condition \eqref{eq:PE eta}, the only stable equilibrium is $\tilde R_{Q_T} = I_3$, all antipodal equilibria are unstable, and the identity is locally exponentially stable. Therefore, $\tilde R_{Q_T} \to I_3$ almost globally, and by extension, $\tilde R \to I_3$ almost globally and locally exponentially.
\end{proof}

\begin{proof}[Proof of Lemma \ref{lemma_uniform_obs}]
        We first compute the state transition matrix of $A(t)$ by writing it as $A(t) = \bar{A} + S(t)$ with  
    $$ S(t) = - I_2 \otimes \bm{\omega}_B ^\times, \qquad \bar{A} = \begin{bmatrix}
        0_{3,3} & I_3 \\ 0_{3,3} & 0_{3,3}
    \end{bmatrix}.$$
    Consider the uniformly invertible change of coordinates $\bar{\bm{x}} = T(t) \bm{x}$ with $T(t) = I_2 \otimes Q_B(t)$.
    Under this transformation, for any  $t \geq 0$, the dynamics reduce to the time-invariant system $\dot{\bar{\bm{x}}} = \bar{A} \bar{\bm{x}}$, which implies that $\bar{\bm{x}}(s) = \bar{\Phi}(s,t)\bar{\bm{x}}(t)$, with $\bar{\Phi}(s,t) = \exp(\bar{A}(s-t))$.
    Therefore, the transition matrix of the original state satisfies $\Phi(s,t) = T(s)^\top \bar{\Phi}(s,t) T(t)$.
    The Gramian of $(A(t), C(t))$ satisfies $W(t, t+ \delta) = T(t)^\top \bar{W}(t, t + \delta) T(t)$, where
    \begin{align} \label{eq:pv_new_gramian}
        \bar{W}(t, t+ \delta) 
        &=  \frac{1}{\delta} \int_t^{t+\delta} \bar{\Phi}^\top(s,t) \bar{C}^\top(s)\bar{C}(s)\bar{\Phi}(s,t) ds,
    \end{align}
    with $\bar{C}(t) = Q_B C(t) T^\top = \left[Q_B\pi_{\bm{y}_{\xi}}Q_B^\top \; 0_{3,3} \right] = [ \pi_{\bm{y}_\xi^{\mathcal{I}}} \; 0_{3,3} ]$. This matrix admits the factorization $\bar{C}(t) = \pi_{\bm{y}_\xi^{\mathcal{I}}(t)} H$, $H := [I_3 \; 0_{3,3}  ]$. Since $\mathrm{rank}\left(\left[\begin{smallmatrix}
        H \\ H \bar{A}
    \end{smallmatrix}\right]\right) = 6$, the $(\bar{A}, H)$ is Kalman observable.
    Thus, under the \emph{p.e.} assumption $\frac{1}{\delta} \int_t^{t+\delta} \pi_{\bm{y}_\xi^{\mathcal{I}}(s)} ds \geq \bar{\mu}I_3$ for all $t \geq 0$, 
    Lemma 2.7 in \cite{hamel2017position} implies that the Gramian \eqref{eq:pv_new_gramian} satisfies $\bar{W}(t, t + \delta) \geq \bar{\mu} I_3$, $\forall t \geq 0$ and the pair $(\bar{A}, \bar{C}(t))$ is uniformly observable. Therefore, the original pair $(A(t), C(t))$ is uniformly observable as well. 
    This in turn guarantees that the unforced error dynamics \eqref{pv error} are globally exponentially stable at the equilibrium $\bm{x} = 0$.
    \end{proof}

\begin{proof}[Proof of Lemma \ref{lemma_pv_yaw_uo}]
To examine uniform observability of $(A_\theta, C_\theta)$, we first define the matrix $A_\theta^\star(t)$, corresponding to the linearization w.r.t. the true relative attitude $R$. One easily verifies that this matrix is
    $$ A_\theta^\star(t) = \begin{bmatrix}
    -\bm{\omega}_B^\times  & I_3 & 0_{3,1} \\ 0_{3,3} & - \bm{\omega}_B^\times & R^\top \bm{a}_T^\times \bm{e}_3 \\ 0_{1,3} & 0_{1,3} & 0
\end{bmatrix} = \begin{bmatrix}
     A(t) & \begin{matrix}
         0_{3,1} \\ R^\top \bm{a}_T^\times \bm{e}_3
     \end{matrix}
     \\ \begin{matrix}
         0_{1,3} & 0_{1,3}
     \end{matrix} & 0
\end{bmatrix}.
$$ 
Uniform observability of $(A_\theta^\star(t), C_\theta(t))$ extends locally to $(A_\theta(t), C_\theta(t))$ by uniform continuity, as established in \citep{hamel2017riccati}.

To simplify the forthcoming computations, we perform the following change of coordinates
$\bar{\bm{x}}_\theta = \bar{T}(t) \bm{x}_\theta$ with 
$$\bar{T}(t) = \begin{bmatrix}
    I_2 \otimes Q_B(t) & 0_{6,1} \\ 0_{1,6} & 1
\end{bmatrix}. $$ 
Under this transformation, the dynamics become
$$ \dot{\bar{\bm{x}}} = \bar{A}_\theta(t) \dot{\bar{\bm{x}}} + O(|\tilde{\theta}|^2), \quad \bar{A}_{\theta}(t) = \begin{bmatrix}
    0_{3,3} & I_3 & 0_{3,1} \\ 0_{3,3} & 0_{3,3} & \bm{b}(t) \\ 0_{1,3} & 0_{1,3} & 0
\end{bmatrix}$$

Let $\bm{\beta}(s,t) = \int_t^s(s-\tau) \bm{b}(\tau) 
d\tau$. The state transition matrix of $\bar{A}_{\theta}(t)$ satisfies $\bar{\bm{x}}_\theta(s) = \bar{\Phi}(s,t) \bar{\bm{x}}_\theta(t)$ and is given by
\begin{equation} \label{eq:phi_bar}
    \bar{\Phi}_\theta(s,t) = \begin{bmatrix}
    I_3 & (s-t)I_3 & \bm{\beta}(s,t) \\ 
    0_{3,3} & I_3 & \int_t^s \bm{b}(\tau) d\tau \\ 
    0_{1,3} & 0_{1,3} & 1
\end{bmatrix}. 
\end{equation}

Then, similarly to the proof of Lemma \ref{lemma_uniform_obs}, we show the observability of $(A_\theta, C_\theta)$ via the observability Gramian
$
        \bar{W}_\theta(t, t+ \delta) 
        =  \frac{1}{\delta} \int_t^{t+\delta} \bar{\Phi}_\theta^\top(s,t) \bar{C}_\theta^\top(s)\bar{C}_\theta(s)\bar{\Phi}_\theta(s,t) ds
    $,
with
\begin{equation} \label{eq:C_bar}
    \bar{C}_\theta(t) = Q_B(t)\left[C, \, 0_{3,1} \right]\bar{T}(t)^\top = \left[\pi_{\bm{y}_{\xi}^{\mathcal{I}}}, \, 0_{3,4} \right]. 
\end{equation}
Using the expression \eqref{eq:phi_bar} of $\bar{\Phi}_\theta(s,t)$, the Gramian takes the form
\begin{align*}
\bar{W}_\theta
    &= \frac{1}{\delta} \int_t^{t+\delta}\begin{bmatrix}
        I_3 \\ (s-t) I_3 \\ \bm{\beta}(s,t)^\top
    \end{bmatrix} \pi_{\bm{y}_{\xi}^{\mathcal{I}}}
    \begin{bmatrix}
        I_3 & (s-t) I_3 & 
        \bm{\beta}(s,t)
    \end{bmatrix} ds.    
\end{align*}
$\bar{W}_\theta$ satisfying \eqref{eq:main Uniform Observability condition} is equivalent to requiring that, for all $t \geq 0$ and all $\bm{\zeta} \in \mathbb{S}^6 \subset \R^7$, $\bm{\zeta}^\top \bar{W}_\theta(t, t+ \delta) \bm{\zeta} \geq \epsilon > 0$. 

Let $\bm{f}(s,t, \bm{\zeta}) = \bar{C}_\theta(t) \bar{\Phi}_\theta(s,t) \bm{\zeta}$. From \eqref{eq:phi_bar} and \eqref{eq:C_bar}, we can write $\bm{f}(t,s,\bm{\zeta}) = \pi_{\bm{y}_{\xi}^{\mathcal{I}}} (\bm{\zeta}_1 + (s-t_k) \bm{\zeta}_2 + \bm{\beta}(s,t_k) \zeta_{3})$.

Following \citep{morin2017uniform}, we argue by contradiction. Suppose that \eqref{eq:main Uniform Observability condition} does not hold. That is, for all $\delta,\mu>0$ 
there exists a sequence $\{t_k\}_{k \in \mathbb{N}}$, and $\bm{\zeta} = (\bm{\zeta}_{1}, \bm{\zeta}_{2}, \zeta_3) \in \mathbb{S}^6$ such that  
\begin{equation}  \label{eq:morin_condition}
    \lim_{k \rightarrow +\infty} \int_{t_k}^{t_k+\bar{\delta}+\delta} \| \bm{f}(t_k,s,\bm{\zeta}) \|^2 ds = 0, 
\end{equation}  
with $\bar{\delta} > 0$ as large as desired.

Using the fact that $\bm{b}(t) = \bm{b}_0 + \bm{b}_1(t)$, we have directly 
$$
\bm{\beta}(s,t) = \frac{1}{2} (s-t)^2 \bm{b}_0 + \int_t^s(s-\tau)
\bm{b}_1(\tau) d\tau.
$$
Accordingly, since the quadratic term in $s$ of $\bm{\beta}(s,t) \zeta_3$ dominates the other terms by a factor $s$, the satisfaction of \eqref{eq:morin_condition} implies that 
they can be neglected  when $\bar{\delta}$ becomes sufficiently large.
Let us now rewrite $\bm{f}$ as $\bm{f}(t,s, \bm{\zeta}) =  \pi_{\bm{y}_{\xi}^{\mathcal{I}}} \bm{\beta}(s,t) \zeta_3$, then
\begin{align*} 
\int_{t}^{t+\bar{\delta}+\delta} \| \pi_{\bm{y}_{\xi}^{\mathcal{I}}} \bm{\beta}(s,t) \zeta_3 \|^2 ds  
&= \int_{t}^{t+\bar{\delta}+\delta} \left\| \pi_{\bm{y}_{\xi}^{\mathcal{I}}} \left( \frac{(s-t)^2}{2}  \bm{b}_0 + \int_t^s(s-\tau)
\bm{b}_1(\tau) d\tau \right) \zeta_3 \right\|^2 ds, \\
&\geq  \int_{0}^{\bar{\delta}+\delta} \frac{1}{8} \| s^2 \zeta_3 \pi_{\bm{y}_{\xi}^{\mathcal{I}}} \bm{b}_0 \|^2ds  - \int_{t}^{t+\bar{\delta}+\delta} \| \zeta_3 \pi_{\bm{y}_{\xi}^{\mathcal{I}}} \int_t^s(s-\tau)
\bm{b}_1(\tau) d\tau \|^2 ds, 
\end{align*}  
Condition \eqref{eq:PE pv} implies $\int_{\bar{\delta}}^{\bar{\delta}+\delta} \|\pi_{\bm{y}_{\xi}^{\mathcal{I}}}  \bm{b}_0 \|^2 ds \geq \delta \mu \|\bm{b}_0\|^2$. Thus,
\begin{align*}
    \int_{0}^{\bar{\delta}+\delta} \| s^2 \pi_{\bm{y}_{\xi}^{\mathcal{I}}} \bm{b}_0 \zeta_3 \|^2 ds 
&\geq  \frac{1}{\delta}|\zeta_3|^2 \int_{0}^{\bar{\delta}+\delta} s^4 ds \int_{0}^{\bar{\delta}+\delta} \| \pi_{\bm{y}_{\xi}^{\mathcal{I}}} \bm{b}_0 \|^2 ds \\
&\geq  \frac{1}{5}(\bar{\delta}+\delta)^5 \mu \|\bm{b}_0 \|^2 |\zeta_3|^2.
\end{align*}
On the other hand, we can write
\begin{align*}
\int_{t}^{t+\bar{\delta}+\delta} \left\| \zeta_3 \int_0^{s}(s -\tau)
\pi_{\bm{y}_{\xi}^{\mathcal{I}}}  \bm{b}_1(\tau) d\tau \right\|^2 ds 
&\leq |\zeta_3|^2 \int_{t}^{t+\bar{\delta}+\delta} 
\left( \int_t^{s} (s-\tau)^2 d\tau \right) \left( \int_t^{s} \| \pi_{\bm{y}_{\xi}^{\mathcal{I}}}  \bm{b}_1(\tau) \|^2 d\tau \right) ds \\
&\leq |\zeta_3|^2 \int_{t}^{t+\bar{\delta}+\delta} 
\frac{(s-t)^3}{3} \left(\int_t^s \| \pi_{\bm{y}_{\xi}^{\mathcal{I}}}  \bm{b}_1(\tau) \|^2 d\tau \right) ds. 
\end{align*} 
Since $\int_t^s \| \pi_{\bm{y}_{\xi}^{\mathcal{I}}}  \bm{b}_1(\tau) \|^2 d\tau \leq \int_t^{t+\delta} \| \pi_{\bm{y}_{\xi}^{\mathcal{I}}}  \bm{b}_1(\tau) \|^2 d\tau \leq \frac{\delta \mu}{4} \|\bm{b}_0\|^2$, one deduces that
\begin{align*}
\int_{t}^{t+\bar{\delta}+\delta} \left\| \zeta_3 \int_0^{s}(s -\tau)
\pi_{\bm{y}_{\xi}^{\mathcal{I}}}  \bm{b}_1(\tau) d\tau \right\|^2 ds 
&\leq \frac{\delta \mu}{12} \|\bm{b}_0\|^2 |\zeta_3|^2 \int_{0}^{\bar{\delta}+\delta} s^3  ds  \\
&\leq \frac{1}{48} (\bar{\delta} + \delta)^4 \delta \mu \|\bm{b}_0\|^2 |\zeta_3|^2.
\end{align*}
Finally, it follows that 
\begin{align*}
    \int_{t}^{t+\bar{\delta}+\delta} \| \bm{f}(t,s,\bm{\zeta}) \|^2 ds  
    &\geq \left( \frac{(\bar{\delta} + \delta)^5}{40}  - \frac{(\bar{\delta} + \delta)^4  \delta}{48} \right) \mu \|\bm{b}_0\|^2 |\zeta_3|^2 \\
    &\geq \frac{1}{40} (\bar{\delta} + \delta)^4\bar{\delta} \mu \|\bm{b}_0\|^2 |\zeta_3|^2.
\end{align*}
Then, the convergence of $ \lim_{k \rightarrow +\infty} \int_{t_k}^{t_k+\bar{\delta}+\delta} \| \bm{f}(t_k,s,\bm{\zeta}) \|^2 ds$ to zero when $\bar{\delta}$ is large implies that $|\zeta_3| \rightarrow 0$.
By the expression of $\bm{f}$, the vanishing of $\zeta_3$ forces all other components of $\bm{\zeta}$ to vanish as well. However, $\bm{\zeta}$ is assumed to be a unit vector, i.e., $|\bm{\zeta}| = 1$, which leads to a contradiction.
One concludes that the pair $(A_\theta, C_\theta)$ is uniformly observable. 

\end{proof}

\end{document}